\documentclass[12pt]{iopart}
\usepackage{cite}
\usepackage{iopams}
\usepackage{epsfig}
\usepackage{subfigure}
\usepackage{graphicx}
\usepackage{url}
\def\eenheid{{\rm 1\kern-0.22em\hbox{\rm l}}}

\DeclareSymbolFont{AMSb}{U}{msb}{m}{n}
\DeclareSymbolFontAlphabet{\Bbb}{AMSb}

\newcommand{\N}{\Bbb{N}}

\newcommand{\R}{\Bbb{R}}
\newcommand{\C}{\Bbb{C}}

\newcommand{\ie}{i.e.\ }

\begin{document}

\today
{\hspace*{\fill} Preprint-KUL-TF-2001/04}

\title{Metastability in the BCS-model}
\author{J.Lauwers\footnote[7]{Bursaal KUL FLOF-10408}, A.Verbeure}

\address{Instituut voor Theoretische Fysica,  
Katholieke Universiteit Leuven,   
Celestijnenlaan 200D,   
B-3001 Leuven, Belgium}
\ead{\mailto{joris.lauwers@fys.kuleuven.ac.be}, \mailto{andre.verbeure@fys.kuleuven.ac.be}}

\begin{abstract}
We discuss metastable  states in the mean-field version of the strong coupling 
BCS-model and study the evolution of a superconducting equilibrium state 
subjected to a dynamical semi-group with Lindblad generator in detailed balance
w.r.t.\ another equilibrium state. The intermediate states are explicitly 
constructed and their stability properties are derived. The notion of
metastability in this genuine quantum system, is expressed by means of energy-entropy 
balance inequalities and canonical coordinates of observables. 
\end{abstract}

\pacs{05.30 
-- 74.20.F 
-- 64.60.M 
-- 42.50.L 
}

\section{Introduction}

One of the most well-known phenomena in physics, is metastability.
Phenomena like supercooled water or hysteresis in magnetic systems are easily
observed and identified with metastability. But in spite of its clear
appearance, the development of a complete theory of metastability  is 
still a hard and unresolved problem, for an overview, see e.g.\ 
\cite{penrose:1987}. 
The basic concept of metastability could be formulated in the following way: a
thermodynamic system is prepared in a special initial state different from the
equilibrium state. If the initial conditions are suitably chosen, the system
will not relax immediately to equilibrium, but it persists a longer period of time
away from the equilibrium state, until some large fluctuation or an external 
disturbance occurs driving the system to equilibrium.  The key problem is 
to formulate an expression for the size of the perturbation characterising the metastable 
regime \cite{ cassandro:1984,neves:1991,penrose:1987, schonmann:1998}.
    
Metastability has intensively been studied for classical models, and
interesting results are obtained for the metastable relaxation in kinetic 
Ising models \cite{cassandro:1984,penrose:1987,neves:1991,schonmann:1998}, we will not 
try to give a full overview of this field but refer to one of these papers 
for more references.  
It was conjectured (see e.g.\ \cite{penrose:1987}) that the metastable evolution is governed by
the growth of droplets of the stable phase in a background of the (initial) 
metastable phase.  Small droplets are probable to disappear again, but bigger 
droplets tend to grow, ultimately driving the system to equilibrium. 
The lifetime of the
metastable phase should then be linked to the probability of creating a
droplet-excitation of critical volume \cite{neves:1991}.
A rigorous result pointing in that direction was obtained in a
paper by Schonmann and Shlosman \cite{schonmann:1998}
where they proved that the `exit-time' could be expressed as a function of the
equilibrium surface-tension of a Wulff-droplet of volume one.
A Wulff-droplet is a typical droplet excitation, for which the shape is found
minimizing the surface terms in the free-energy.
This activity and success for classical models is in sharp contrast with the situation for
quantum mechanical models where little is known. 
In a recent paper \cite{nachtergaele:2001}, droplet states were constructed 
for the quantum mechanical XXZ-Heisenberg model. 

In this note, we want to develop ideas of \cite{schonmann:1998} in order to
approach the phenomenon of metastability for quantum systems. 
We put forward that the underlying concept for exit-times and the metastable 
evolution is situated in the behaviour of non-extensive terms such as fluctuations 
of relevant observables, 
like the non-extensive (equilibrium) surface-tension of droplet excitations 
determines the metastable evolution in Ising models \cite{schonmann:1998}.
For the BCS-model \cite{thirring:1967,thirring:1968} studied in this paper, 
we find a characterisation for metastability and define exit-times for 
different observables, expressed as a
function of the equilibrium expectation values of the `normal coordinates' of the
observable under consideration.
The metastable evolution at an arbitrary temperature between two
extremal superconducting phases is studied. The evolution is driven by a
semigroup with Lindblad generator \cite{alicki:1976,kossakowski:1977}.
In fact we consider a detailed balance dynamics between two different fixed
phase states and realise the evolution from one equilibrium state to
another equilibrium state with a different phase. 
The intermediate states are
explicitly constructed and their thermodynamic properties are derived.
These states are not invariant under the Hamiltonian evolution, but
the correlation inequalities (section \ref{sectcor}) are satisfied for the
normal coordinates of observables. 
We present a simple criterion to distinguish observables exhibiting monotone
relaxation or metastable relaxation, and in the latter case, an expression 
for the exit-time is given. The exit-time is the time necessary to leave 
the initial state on the basis of having reached the maximum value of the 
observable under consideration. 
In the BCS-model, the observables which are invariant under the gauge transformation
group of the broken symmetry \cite{goderis:1991,michoel:2001,lauwers:2001}, come over as 
relevant observables. They all exhibit metastable relaxation and all have the same
exit-time.

\section{The BCS-model}\label{sectBCS}

The strong-coupling BCS-model is described by the local Hamiltonians
\cite{thirring:1967,thirring:1968}
\begin{equation}\label{BCSHAM}
H_N = - \frac{1}{N}\sum_{i\ne j = 1}^{N}\sigma_i^+\sigma_j^- + \epsilon
\sum_{i=1}^{N}\sigma_i^z,\qquad \qquad \epsilon > 0
\end{equation}
where $\sigma^+$, $\sigma^-$ and $\sigma^z = \sigma^+\sigma^- - \sigma^-\sigma^+$ 
are the well-known Pauli matrices. $H_N$ acts on the Hilbert space 
$\bigotimes_{i= 1}^N \C_i^2$. 

The equilibrium states $\omega_\beta$ studied here are elements of the set of 
$(\tau_t,\beta)$-KMS states in the thermodynamic limit $(N \uparrow
\infty)$ of this model, \ie the states satisfying the equilibrium conditions of
Kubo, Martin and Schwinger \cite{bratteli:1996}:
\begin{equation}\label{kms}
\omega_\beta(BA) = \omega_\beta(A\tau_{i\beta}B) \qquad \forall A,B \in
\mathcal{B},
\end{equation}
on the infinite tensor product algebra $\mathcal{B} = \bigotimes_{i= 1}^\infty M_2$
at an inverse temperature $\beta$ and with the reversible Heisenberg dynamics: 
$$\tau_t(.) = w-\lim_{N \to \infty} = e^{itH_N} .\, e^{-itH_N}.$$
The extremal points of the set of equilibrium states are given by the symmetric product states
\cite{fannes:1980}:
\begin{equation}\label{prod}
\omega_\lambda(.) =\prod_{i=1}^\infty \tr \rho_{\lambda}\, .
\end{equation}
on the infinite tensor product algebra $\mathcal{B}= \bigotimes_{i= 1}^\infty M_2$;
$\rho_\lambda $ is a $2 \times 2$ density matrix, given by the solutions of the
gap-equation:
\begin{equation}\label{gap1}
\rho_\lambda = \frac{\exp[-\beta h_{\lambda}]}{\tr \exp[-\beta
     h_{\lambda}]},
\end{equation}
where
   \begin{equation}\label{heff}
     h_{\lambda} = \epsilon\sigma^z - \lambda \sigma^- - \bar{\lambda} \sigma^+,
   \end{equation}
and with order parameter $\lambda$, given by,
 \begin{equation}\label{op} 
     \lambda  =  \tr \rho_\lambda (\sigma^+) =  \omega_\lambda
     (\sigma^+). 
 \end{equation}
This equation can be transformed into the equation
      \begin{equation}\label{gap2}
      \lambda (1 - \frac{1}{2k}\tanh[\beta k]) = 0,  
      \end{equation}
where $k = \sqrt{\epsilon^2 + |\lambda|^2}$; $\{-k,k\}$ is the spectrum of the
effective Hamiltonians $h_\lambda$ (\ref{heff}), which is independent of the 
phase of the order parameter $\lambda \in \C$.
It can readily be seen that this equation (\ref{gap2}) has always a solution
$\lambda =0$. This corresponds to the normal phase state. 
Solutions with $\lambda \ne 0$
exist if the following conditions are satisfied:
\begin{equation}\label{crit}
\left \{ 
   \begin{array}{lcl} 
     \epsilon & < & 1/2, \\
     \beta & > & \beta_c = 
     \frac{1}{2\epsilon}\log\left(\frac{1+2\epsilon}{1-2\epsilon}\right).
   \end{array} 
   \right .
\end{equation} 
These solutions $\lambda \ne 0$, are understood to describe
the superconducting phase states. The inverse temperature $ \beta  > \beta_c$,
fixes only $|\lambda|$, the norm of the order parameter. 
The phase of the order parameter $\phi$, defined by $\lambda = |\lambda|e^{i\phi}$, $\phi 
\in [0, 2\pi]$, remains free to choose. 
This leads to an infinite 
degeneracy of the states in the superconductive regime and is due to spontaneous
symmetry breaking \cite{goderis:1991,michoel:2001}. 
As this phase becomes important in the remainder of this article, it will 
from now on explicitly be labelled, \ie we denote the order parameter (\ref{op})
as $\lambda e^{i\phi}$, with a phase $\phi \in [0, 2\pi]$, and norm $\lambda 
\in \R^+$. 
Furthermore, we fix now a certain subcritical temperature $\beta > \beta_c$, 
and hence the norm of the order parameter, which will now be labelled 
by $\lambda \in \R^+$. The superconducting pure phase states (\ref{gap1}) are 
therefore from now on denoted by $\omega_\phi$ instead of
$\omega_{|\lambda|e^{i\phi}}$. 

The mechanism of spontaneous symmetry breaking can be explicitly seen as
follows: The Hamiltonian $H_N$ 
(\ref{BCSHAM}) is invariant under the norm-continuous gauge transformation 
automorphism group  $\mathcal{G}=\{ \alpha_\psi\,|\, \psi \in [0,2\pi]\}$ 
on $\mathcal{B}$, defined by the action:
\begin{equation}\label{ijkgr}
\alpha_{\psi}(\sigma^+_i) = e^{i\psi}\sigma^+_i.
\end{equation}
On finite subsets $\Lambda \subset \N$, these transformations $\alpha_\psi$ are implemented by 
the unitaries
\begin{equation*}
U_\Lambda^\psi = e^{i\psi Q_\Lambda/2}
\end{equation*}
where $Q_\Lambda$ is the (local) infinitesimal generator: 
\begin{equation*}\label{ijkq}
Q_\Lambda = \sum_{i \in \Lambda} \sigma^z_i.
\end{equation*}
Clearly $\alpha_\psi(H_N)= U_{\Lambda}^{\psi} H_N U_{\Lambda}^{\psi\ *} = H_N$
with $ \{1,\ldots,N\} \subset \Lambda$, 
but this symmetry is broken in the equilibrium states of the 
superconducting phase, we have:
\begin{equation}\label{ijkbroken}
 \omega_\phi \left ( \alpha_{\psi}( \sigma^+_i )\right ) = 
 \lambda e^{i(\phi+\psi)}
  \ne  
  \lambda e^{i\phi} 
 = \omega_\phi (\sigma^+_i).
\end{equation}
In fact the gauge group $\mathcal{G}$ establishes a relation between the 
superconducting pure phase states with different phase-factors:
\begin{equation*}
\omega_\phi(\alpha_\psi(X)) = \omega_{\phi + \psi}(X), \quad \forall X \in
\mathcal{B},\ {\rm and}\ \phi,\psi \in [0, 2\pi].
\end{equation*}
In the following  we  learn that the observables $X \in
\mathcal{B}$ which are invariant under the gauge transformations, \ie 
$\alpha_\psi(X) = X$, are relevant in order to determine the
metastable evolution between two superconducting states.

We conclude this paragraph by remarking  that a common method to 
single out an equilibrium state with fixed phase $\phi$, consists of
adding a thermodynamic unimportant term to the 
local Hamiltonians (\ref{BCSHAM}), such as a vanishing external field:
\begin{equation}\label{transfield}
-\frac{1}{N}\sum_{i=1}^N \sigma_i^+ e^{-i\phi} + \sigma_i^- e^{i\phi}, 
\end{equation}
forcing the system in the limiting Gibbs state $\omega_\phi$  (\ref{gap1}) 
of the superconducting phase. This fixes the phase to $\phi$.
 
\section{Thermodynamic stability}\label{sectcor}
An alternative way to characterise equilibrium states, is given by the
correlation inequalities\cite{fannes:1978}. These represent conditions, which have been proven to be
equivalent to the $\beta$-KMS condition (\ref{kms}), while their interpretation is related to
the principle of minimum free energy. In other words, they are an expression for the 
thermodynamic stability of the  KMS (equilibrium) states
\cite{bratteli:1996}:
\newline
{\bf Energy-Entropy Balance Inequalities}
\newline
{\em
Let $\tau_t(.) = \lim_N e^{itH_N}.\,e^{-itH_N}$ be the Heisenberg dynamics,
(defined in a weak limit sense) and $\beta$ the inverse temperature.
A state $\omega$ is a $(\tau_t,\beta)$-KMS state if and only if for all $X
\in Dom(\delta)$ 
\begin{equation}\label{cor}
-i \beta \omega(X^*\delta X) \geq
\omega(X^*X)\log\frac{\omega(X^*X)}{\omega(XX^*)},
\end{equation}
 with $\delta(.)$ the infinitesimal generator of the dynamics ($\tau_t$).  
}

 In this paper we concentrate on the BCS-model.
Since this model is of mean-field type, we only need to consider product states
(see section \ref{sectBCS}).
For such states the correlation inequalities for operators $X \in \mathcal{B}$
of the form $X = X_1\otimes X_2 \otimes \ldots \otimes X_n$ with 
$X_1,X_2,\ldots, X_n \in M_2$
follow from the correlation inequalities for $X_1,X_2,\ldots X_n$. 
Therefore it is sufficient to consider only one-point operators, and (\ref{cor})
is reduced to:  
\newline
{\bf Energy-Entropy Balance Inequalities in the BCS-model}
\newline
{\em
A product state $\rho$ on $\mathcal{B} = \bigotimes_{i = 1}^\infty M_2$ is a 
$(\tau_t,\beta)$-KMS state for the BCS-model (\ref{BCSHAM})
if and only if
for every one-point operator $X \in M_2$ the following inequality holds:
\begin{equation}\label{corp}
 \beta \rho(X^*[h_\rho,X]) \geq
 \rho(X^*X)\log\frac{\rho(X^*X)}{\rho(XX^*)},
\end{equation}
with $h_\rho$ the effective Hamiltonian in this state, \ie $h_\rho =
\epsilon\sigma^z -\rho(\sigma^+)\sigma^- -\rho(\sigma^-)\sigma^+$.
}

The interpretation of these inequalities is the following: the l.h.s. of
(\ref{cor}) or (\ref{corp}) reflects the change in energy if we alter the state with a
`perturbation $X$'.  
The r.h.s. of inequalities (\ref{cor}) or (\ref{corp}) originates from the
corresponding change in entropy $S(\rho) = - \tr \rho \log\rho$ under the
`perturbation $X$'.
For example, if the criterion  (\ref{corp}) is applied to 
unitaries $U \in M_2$, with $U^*U = \mathbf{1} = UU^*$, 
and we substitute $U$ for $X$ in (\ref{corp}), it 
is reduced to:
\begin{equation*}
\rho(U^*[h_\rho,U]) \geq 0.
\end{equation*}  
This inequality expresses the fact that the 
the local state $\tr(\rho\,.)$ must have a lower internal energy than the perturbed
states  $\tr( U \rho U^*\,.)$,
while the entropy of these states remain unchanged: $S(\rho) = S(U \rho U^*)$.

If a state $\omega$ satisfies condition (\ref{cor}) for an operator 
$X$, we say that $\omega$ is stable under the `perturbation $X$'. 

\section{The perturbed states}\label{sectPS}
In classical kinetic models \cite{penrose:1987,cassandro:1984,
neves:1991,schonmann:1998} 
the metastable evolution is introduced by a dynamical semigroup of 
dissipative maps (e.g.\ Glauber dynamics), satisfying the detailed balance conditions 
w.r.t.\ the asymptotic equilibrium measure. The notion of detailed balance and
dissipative evolutions are generalised for quantum systems
\cite{alicki:1976,kossakowski:1977}. 
The quantum dynamical semigroups as well as the classical Glauber dynamics do
share the same physical background, in the sense that they can be constructed as
the result of a weak coupling of the system with a temperature reservoir system
\cite{davies:1976}. 

A continuous 1-parametergroup $\{\gamma(t)|\,t \geq 0 \}$ of linear maps 
$\gamma(t)$ on the 
algebra $\mathcal{B}= \bigotimes_{i=1}^{\infty} M_2$, is called a dynamical semigroup if 
for every $t \geq 0$, $\gamma(t)$ is a completely positive, unity preserving
map on $\mathcal{B}$ and $\gamma(0)$ is the identity map.

Let $L$ be the (densely defined) infinitesimal generator of such a dynamical 
semigroup, \ie
\begin{equation*}
\gamma(t) = e^{tL},\quad \forall t \geq 0.
\end{equation*}
The dynamical semigroup is then called dissipative if the generator $L$ is
self-adjoint, \ie $L(A^*) = L(A)^*$  $\forall A \in \mathcal{B}$, and
satisfies the following inequalities:
\begin{equation}\label{diss}
L(A^*A) \geq A^*L(A) + L(A^*)A, \quad \forall A \in \mathcal{B}.
\end{equation} 
Let $\omega$ be a state on $\mathcal{B}$, the dynamical semigroup 
$(\gamma(t))_{t \geq 0}$ is said to satisfy the detailed balance conditions
w.r.t.\ the state $\omega$ if the following duality property holds:
\begin{equation*}
\omega(XL(Y)) = \omega(L(X)Y)
,\quad \forall X,Y \in \mathcal{B}.
\end{equation*} 
Consider now one of the extremal superconducting phase states $\omega_\phi$ 
(\ref{gap1}), we can construct a dynamical semigroup $ \{ 
\gamma_\phi(t)|\, t\geq 0\}$ with generator $L_\phi$, satisfying the 
condition of detailed balance w.r.t.\ this state. 
Because we are dealing with product states, the generator $L_\phi$ of 
$(\gamma_\phi(t))$ is globally defined if it is defined on the local sites, 
\ie
\begin{equation*}
L_\phi: M_2 \to M_2.
\end{equation*}
It is given by \cite{alicki:1976,kossakowski:1977}:
\begin{equation}\label{L}
L_\phi(.) = \sum_{i,j} \exp[-\beta(\epsilon_i
-\epsilon_j)/2]\left(E_{ij}^*[\,.,E_{ij}] + [E_{ij}^*,.\,]E_{ij} \right),
\end{equation}
where $E_{ij} = |\psi_i\rangle\langle \psi_j|$ stands for the matrix units in 
the base of eigenvectors \mbox{$\{|\psi_i\rangle|\, i \in \{-,+\}\}$} of 
$h_\phi$, the effective Hamiltonian (\ref{heff}) of the $(\tau_t,\beta)$-KMS 
state $\omega_\phi$ (\ref{gap1}); 
$|\psi_i\rangle $ is the eigenvector corresponding to the eigenvalue 
$\epsilon_i$ of $h_\phi$.
It is immediately checked that $L_\phi$ is dissipative (\ref{diss}) and 
satisfies the detailed balance conditions in the state $\omega_\phi$ with fixed
phase $\phi$, \ie
\begin{equation}\label{DB2}
\omega_\phi(X L_\phi(Y)) = \omega_\phi(L_\phi(X) Y) \qquad \forall X,Y \in
M_2.
\end{equation}
The action of this generator (\ref{L}) is naturally extended to operators 
$X \in \mathcal{B}$ of the form  $X = X_1\otimes X_2 \otimes \ldots \otimes 
X_n$ with $X_1,X_2,\ldots, X_n \in M_2$ by
\begin{equation*}
L_\phi( X_1\otimes X_2 \otimes \ldots \otimes X_n) = 
\sum_{i = 1}^n  X_1 \otimes \ldots \otimes
L_\phi( X_i)\otimes \ldots \otimes  X_n.
\end{equation*}
Since $\omega_\phi$ is a symmetric product state (\ref{prod}),
the dissipativity and the detailed balance properties (\ref{DB2}) are
preserved for these more general operators and follow from 
the properties of their one-site factors.  
The dynamical semigroup $(\gamma_\phi(t))_{t\geq0}$ is then defined by:
\begin{equation}\label{gamma}
\gamma_\phi(t) = e^{tL_\phi},\quad \forall t\geq 0. 
\end{equation} 
Remark that the detailed balance properties (\ref{DB2}) guarantee 
that $\omega_\phi$ is stationary under this dynamical semigroup, 
while any other $(\tau_t,\beta)$-KMS state with a different phase-factor 
is not invariant. 

Suppose now that our system is prepared in a $\beta$-KMS state
$\omega_{-\phi}$ at time $t =0$. Based on ideas from classical models
\cite{penrose:1987,cassandro:1984,neves:1991,schonmann:1998},
we apply at $t \geq 0$, an evolution $(\gamma_\phi(t))$ (\ref{gamma}) with  
phase $\phi$, and the system is forced to evolve accordingly. 
This implies that the system will ultimately relax to the equilibrium state 
$\omega_{\phi}$. The intermediate states $\omega_t$ are introduced by:
\begin{equation}\label{omegat}
\omega_t(.) = \omega_{-\phi}(e^{tL_\phi}.\,),
\end{equation}
where the intermediate states $\omega_t$ are constructed by applying the 
dynamical semigroup $(\gamma_\phi(t))_{t\geq0}$ (\ref{gamma}) with phase $\phi$ to the
initial equilibrium state $\omega_{-\phi}$ with phase $-\phi$. 
These states $\omega_t$ are, by construction, again product states, and their
density matrices can explicitly be calculated.
They are understood to describe a metastable regime. For $t$ small enough, the state
$\omega_t$ will still be close to the initial state $\omega_{-\phi}$, while if $t$
becomes large enough, the system will be relaxing to the equilibrium state 
$\omega_{\phi}$, where the words ``close'' and ``almost'' have to get a precise
meaning. 

We proceed now with the explicit construction of the metastable states
$\omega_t$ (\ref{omegat}).
The spectral decomposition of the generator $L_{\phi}$ (\ref{L}) 
and the dissipative evolution $\gamma_\phi(t)$ is given in terms 
of the matrix units $(E_{ij}^\phi)$ of the corresponding asymptotic effective 
Hamiltonian, which can be written as 
$h_\phi = k E^\phi_{++} -k E^\phi_{--}$. We compute that: 
\begin{equation}\label{lbasis}
\fl L_\phi(\mathbf{1}) = 0;\quad
    L_\phi(D)  =  - dD;\quad
    L_\phi(E^\phi_{+-})  =  -cE^\phi_{+-};\quad 
    L_\phi(E^\phi_{-+})  =  -cE^\phi_{-+},
 \end{equation}
where $d = 4 \cosh(\beta
k), c = 2 + 2 \cosh(\beta k) $ and $D = e^{\beta k} E^\phi_{++} - e^{-\beta k} E^\phi_{--}$. 
Hence, the expectation values for the operators (\ref{lbasis}) in the state
$\omega_t$ (\ref{omegat}) are given by: 
\begin{equation}\label{lbasisinomegat}
\eqalign{
   \omega_t(\mathbf{1}) = 1;\\
   \omega_t(D)  =  e^{-td}\omega_{-\phi}(D);\\
   \omega_t(E^\phi_{+-})  =  e^{-tc}\omega_{-\phi}(E^\phi_{+-});\\ 
   \omega_t(E^\phi_{-+})  =  e^{-tc}\omega_{-\phi}(E^\phi_{-+}).
}
\end{equation}
From these equations (\ref{lbasisinomegat}) and linearity, all expectation values in
$\omega_t$, and hence the density matrix $\rho_t$ at time $t$, 
can be calculated. 

\subsection{Exit-times and normal coordinates}

Remark that any observable $X \in M_2^{sa}$ can be developed in its normal
coordinates for the asymptotic equilibrium state $\omega_{\phi}$ (\ref{DB2}), 
\ie
\begin{equation}\label{xinnc}
X = \omega_{\phi}(X)\mathbf{1} + a_\phi^+(X) + a_\phi^-(X)+ a^0_{\phi}(X). 
\end{equation} 
The operators $a_\phi^+(X)$ and $a_\phi^-(X)$ are understood to be the creation,
resp.\ annihilation operator of the normal modes determined by $X$, they are
given in terms of $X$ and the projection operators $E_{++}^\phi$ and 
$E_{--}^\phi$ on the eigenspaces of the asymptotic effective 
Hamiltonian $h_\phi$, corresponding to positive, resp.\ negative energy:
\begin{eqnarray}
a_\phi^+(X) & = & E^\phi_{++}XE^\phi_{--}; \label{a+}\\
a_\phi^-(X) & = & E^\phi_{--}XE^\phi_{++}, \label{a-}
\end{eqnarray}
while $a_\phi^0(X)$ is the canonical constant of motion determined by $X$:
\begin{equation}\label{a0}
a^{0}_\phi(X) = E^\phi_{++}XE^\phi_{++} + E^\phi_{--}XE^\phi_{--} -
\omega_{\phi}(X)\mathbf{1}.
\end{equation}
Indeed, $a_\phi^0(X)$ is a constant of motion in the state $\omega_\phi$, since
it commutes with the effective Hamiltonian $h_\phi$ and:
\begin{equation*}
\frac{{\rm d}}{{\rm d}t} \omega_\phi(A\tau_t(a_\phi^0(X))B)  
 =i\omega_\phi(A[h_\phi,\tau_t(a_\phi^0(X))]B) = 0, \quad \forall A,B,X \in M_2.
\end{equation*}
Note that these operators (\ref{a+}, \ref{a-}, \ref{a0})  all have expectation
value zero in the asymptotic state $\omega_\phi$. They are an expression
for the fluctuations of $X$ around its asymptotic equilibrium value 
$\omega_\phi(X)$ (see also below). 
Using (\ref{lbasisinomegat}) and (\ref{xinnc}) we get an expression for the 
expectation value of $X$ in $\omega_t$:
\begin{equation}\label{omegatx}
\omega_t(X) = \omega_\phi(X) + \omega_{-\phi}\bigg(a^+_\phi(X)
+ a^-_\phi(X)\bigg)e^{-tc} +
\omega_{-\phi}\bigg(a^{0}_\phi(X)\bigg)e^{-td}.
\end{equation}
The time limits $t \to 0$ and $t \to \infty$ of expression 
(\ref{omegatx}) yield clearly the equilibrium expectation values 
$\omega_{-\phi}(X)$ resp.\ $\omega_{\phi}(X)$: 
\begin{eqnarray*}
\lim_{t \to 0} \omega_t(X) &=& \omega_{-\phi}(X); \\
\lim_{t \to \infty} \omega_t(X) &=& \omega_{\phi}(X),
\end{eqnarray*}
\ie this  time evolution describes the transition between the two states under
consideration, \ie the transition from $\omega_{-\phi} $ to $ \omega_{\phi}$.

In general, the evolution of $\omega_t(X)$  (\ref{omegatx}) can express two types
of behaviour as a function of the time $t$. This is easily derived from the 
analysis the functions 
\begin{equation}\label{fxt}
 f_X(t):\R^+ \to \R : t \mapsto \omega_t(X),\quad \forall X \in M_2^{sa},
\end{equation}
 and their time derivatives.

Let us illustrate this with two pictures (Fig. \ref{fig1}):
\begin{figure}[h]
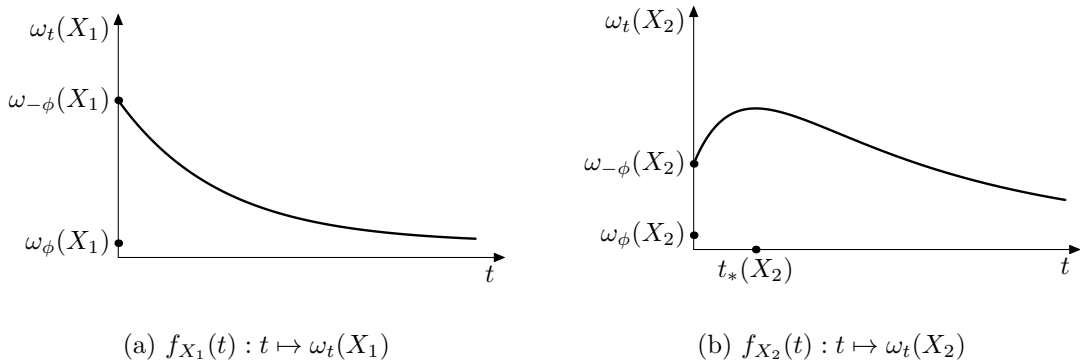
\label{fig1}
\centering
\mbox{\subfigure[$f_{X_1}(t): t \mapsto \omega_t(X_1)$]{\includegraphics{fig-1.1}}
\label{figX1}
\qquad
\subfigure[$ f_{X_2}(t) :t \mapsto \omega_t(X_2) $]{\includegraphics{fig-2.2}}
\label{figX2}
}
\caption{ Typical pictures of monotone (a) and metastable (b) relaxation}
\end{figure}

Firstly, $\omega_t(X)$ can relax monotonically to its 
asymptotic value $\omega_\phi(X)$, and thus behaves qualitatively as shown in figure
\ref{fig1}(a).
This behaviour is met if the function $f_X(t)$ (\ref{fxt}) has only an
extremum at $t =0$. Calculating the time derivative of $f_X(t)$ (\ref{fxt}), 
we see that this happens whenever 
\begin{equation}\label{monotone}
-\frac{\omega_{-\phi}(a_\phi^0(X))}
{\omega_{-\phi}(a_\phi^+(X)) +\omega_{-\phi}(a_\phi^-(X)) } \leq \frac{c}{d},
\end{equation}
where
\begin{equation*}
\frac{c}{d} = \frac{1 + \cosh(\beta k)}{2\cosh(\beta k)}.
\end{equation*}
This condition is satisfied if e.g.\ the second and the third term on the
r.h.s.\ of expression (\ref{omegatx}) have the same sign.
The relaxation is exponentially fast and is determined by the constants 
$c = 2 + 2\cosh(\beta k)$ and $d = 4\cosh(\beta k)$ (\ref{lbasis}).

The second possibility is that condition (\ref{monotone}) is violated, \ie if 
\begin{equation}\label{metastable}
-\frac{\omega_{-\phi}(a_\phi^0(X))}
{\omega_{-\phi}(a_\phi^+(X)) +\omega_{-\phi}(a_\phi^-(X)) } 
> \frac{1 + \cosh(\beta k)}{2\cosh(\beta k)}.
\end{equation}
 In this case the function $f_X(t):t \mapsto \omega_t(X)$ (\ref{fxt}) 
has an extremum, reached at a time $t_*(X) > 0$, which may 
depend on the observable $X$.
A typical picture of such a behaviour is shown if figure \ref{fig1}(b).

The interpretation of this extremum is the following. The time $t_*(X)$ is a
time-scale which indicates where the expectation value $\omega_t(X)$ leaves the
metastable regime and the relaxation to the asymptotic equilibrium value
$\omega_\phi(X)$ starts. For $t < t_*(X)$, $\omega_t(X)$ moves away from the
original equilibrium value $\omega_{-\phi}(X)$, this behaviour with an 
increasing distance from equilibrium corresponds to the
metastable regime. At time $t = t_*(X)$, this evolution comes to an end since
$\omega_{t}(X)$ is now at its maximum distance from the equilibrium value 
$\omega_{-\phi}(X)$.
For $t > t_*(X)$, we see an other type of behaviour. The system is again
approaching equilibrium, but $\omega_t(X)$ is not returning to the  initial
equilibrium value $\omega_{-\phi}(X)$, it relaxes now 
(monotonically) to its new equilibrium value $\omega_{\phi}(X)$.

Since the time $t_*(X)$ marks where $\omega_t(X)$ leaves metastability and the
relaxation to the asymptotic value is started, we call $t_*(X)$ the exit-time
for the observable $X$.
The exit-time $t_*(X)$ can be found 
calculating the extrema of function (\ref{fxt}), \ie  $t_*(X)$ is the 
time $t$ for which $\left. \frac{{\rm d}}{{\rm d}t}f_X(t)\right|_{t} =0$, 
it is readily computed being:
\begin{equation}\label{tsterx}
t_*(X) = \frac{1}{d -c}\left( \log\frac{d}{c} + 
\log \left|\frac{\omega_{-\phi}(a_\phi^0(X))}
{\omega_{-\phi}(a_\phi^+(X)) +\omega_{-\phi}(a_\phi^-(X)) }
\right| 
\right).
\end{equation} 
In this expression, it is clear that the exit-time for the expectation value of
an observable $t_*(X)$ is
determined by the constants $c,d$ (\ref{lbasis}) and the ratio between the
initial-state expectation values of the normal coordinates of $X$
(\ref{a+}), (\ref{a-}) and (\ref{a0}).

In general we can formulate the following statement:
\newline
{\bf Metastable Relaxation for the BCS-model}
\newline 
{\em
A system prepared in a initial product state $\omega_{-\phi}$ on $\mathcal{B}
= \bigotimes_{i=1}^\infty M_2$
will relax to a superconducting equilibrium state $\omega_\phi$ (\ref{gap1}) at
subcritical temperature $\beta$ (\ref{crit}) according to the evolution induced
by a dynamical semigroup (\ref{gamma}) with a dissipative generator $L_\phi$
(\ref{L}),
satisfying the quantum detailed balance conditions in state $\omega_\phi$.
The behaviour of this relaxation depends on the chosen observables 
$X \in M^{sa}_2$:  
\begin{itemize}
\item {\em Monotone Relaxation} 

The relaxation of the expectation value of an observable $X \in M^{sa}_2$ is 
monotone if the following holds:
\begin{equation*}
-\frac{\omega_{-\phi}(a_\phi^0(X))}
{\omega_{-\phi}(a_\phi^+(X)) +\omega_{-\phi}(a_\phi^-(X)) } \leq 
\frac{1 + \cosh(\beta k)}{2\cosh(\beta k)},
\end{equation*}
where $a^+_\phi(X),\ a^-_\phi(X)$ and $a^{0}_\phi(X)$ are the normal 
coordinates of $X$ in
the equilibrium state $\omega_\phi$ as defined by equation (\ref{xinnc}), and 
$k = \sqrt{\epsilon^2 + \lambda^2}$, the positive eigenvalue of the effective
Hamiltonians at inverse temperature $\beta$ (\ref{heff}).
\item {\em Metastable Relaxation}

The expectation value of an observable $X \in M^{sa}_2$ relaxes metastably if the 
following is true:
\begin{equation*}
-\frac{\omega_{-\phi}(a_\phi^0(X))}
{\omega_{-\phi}(a_\phi^+(X)) +\omega_{-\phi}(a_\phi^-(X)) } 
> \frac{1 + \cosh(\beta k)}{2\cosh(\beta k)}.
\end{equation*}
In this case we define the {\sl exit-time} for the observable $X$, $t_*(X)$ as the time when
the expectation value of $X$ has past its extremal value and the relaxation to
equilibrium  starts. $t_*(X)$ is then given by the expression:
\begin{equation*}
\fl
t_*(X) = \frac{1}{2\cosh(\beta k) - 2}\left( \log\frac{2\cosh(\beta k)}{1 + 
\cosh(\beta k)} + \log \left|\frac{\omega_{-\phi}(a_\phi^0(X))}
{\omega_{-\phi}(a_\phi^+(X)) +\omega_{-\phi}(a_\phi^-(X)) }
\right| 
\right).
\end{equation*}
\end{itemize}}

An interesting point of these considerations is that the decomposition
(\ref{xinnc}) and the construction of the exit-times is in terms of operators
which are closely related to quantum fluctuations. Quantum fluctuations
$F_\omega(X)$ 
are the limits $n \to \infty$ of operators $F_n(X)$ defined by:
\begin{equation*}
F_n(X) = \frac{1}{\sqrt{n}}\sum_{i=1}^n X_i - \omega(X),
\end{equation*}
where $X$ is a local observable and $X_i$ is a copy of $X$, translated to site
$i$.
In \cite{goderis:1989,goderis:1990} one has proved the existence of the limit:
\begin{equation*}
\lim_{n\to \infty} \omega\left(F_n(X)^2\right) =
\tilde{\omega}\left(F_\omega(X)^2\right),
\end{equation*}
defining a dynamical system on the level of the algebra of fluctuations
$\{F_\omega |\, X \}$, and defining a state $\tilde{\omega}$ on the algebra of
fluctuations. Also a dynamics $(\tilde{\tau}_t)$ of fluctuations induced by
the original one $(\tau_t)$, is defined by the formula
\begin{equation}
\tilde{\tau}_t F_\omega(X) = F_\omega(\tau_tX).
\end{equation}
In \cite{goderis:1989,goderis:1990} it is proved this dynamical system is 
always a bosonic system. This is worked out in great detail. 
We refer to \cite{goderis:1989,goderis:1990} for more details and more
precise information on this subject.
The point is that the exit-times (\ref{tsterx}) can formally be expressed in term of
fluctuations, as follows:
\begin{equation*}
t_*(X) = \frac{1}{d -c}\left( \log\frac{d}{c} + 
\log \left|\frac{\tilde{\omega}_{-\phi}(F_\phi(X_0)) }
{\tilde{\omega}_{-\phi}(F_\phi(Q_X))}\right| 
\right),
\end{equation*}
where
\begin{eqnarray*}
F_\phi(Q_X) &=& \lim_{n \to \infty} \frac{1}{\sqrt{n}}\sum_{i=1}^n (a_\phi^+(X)
+ a_\phi^-(X))_i; 
\\
F_\phi(X_0) &=& \lim_{n \to \infty} \frac{1}{\sqrt{n}}\sum_{i=1}^n
(a_\phi^0(X))_i, 
\end{eqnarray*} 
and $\tilde{\omega}_{-\phi}$ is the limiting state with phase $-\phi$
on the fluctuation algebra. The thermodynamic limit in
nominator and denominator is taken jointly. 
Although the mathematical formulation of this computation is far from completely
coherent with the general theory of quantum fluctuations 
\cite{goderis:1989,goderis:1990}, we are tempted to conjecture that this 
construction might be the key to the understanding of metastability in a 
broader class of quantum systems.  
\subsection{Invariant observables under the gauge transformation group}
Let us now continue the study of the metastable relaxation in the BCS-model 
between two equilibrium states, in particular the evolution from 
$\omega_{-\phi}$ to $\omega_{\phi}$ for gauge invariant observables.  
As explained, these superconducting states are not
invariant under the gauge-symmetry $\alpha_\psi,\ \psi \in [0,2\pi]$ 
(\ref{ijkgr}) of the BCS-Hamiltonians (\ref{BCSHAM}), in particular:
\begin{equation*}
\omega_{\phi} = \omega_{-\phi}\circ\alpha_{2\phi}.
\end{equation*}
Observables $X \in M_2$ which are invariant under this gauge transformation 
group $\mathcal{G}$ (\ref{ijkgr}) \ie satisfying
\begin{equation}\label{xinvijk}
\alpha_\psi(X) = X \ {\rm or}\quad [\sigma^z,X] =0,
\end{equation}
have the same expectation value in all superconducting phase states, hence 
\begin{equation*}
\omega_{-\phi}(X) = \omega_{\phi}(X).
\end{equation*} 
Developing the observable $X$ in normal coordinates w.r.t.\ $\omega_\phi$ (\ref{xinnc})
yields
\begin{equation*}
\omega_{-\phi}\left(a^{+}_\phi(X)\right) +
\omega_{-\phi}\left(a^{-}_\phi(X)\right) +
\omega_{-\phi}\left(a^{0}_\phi(X)\right) = 0.
\end{equation*}
Hence we can rewrite equation (\ref{omegatx}) yielding:
\begin{equation*}
 \omega_t(X) = \omega_\phi(X) - \omega_{-\phi}\left(a^0_\phi(X)\right)
 \left( e^{-tc} - e^{-td}\right).
\end{equation*}
This implies that the relaxation for observables invariant under 
the gauge transformation group $\mathcal{G}$ (\ref{ijkgr}) is of the metastable type. The transition between the metastable regime and the relaxation regime happens at
the same moment \ie the invariant observables $\alpha_\psi(X) = X$ share the same exit-time (\ref{tsterx}):
\begin{equation}\label{t*}
t_* = \frac{\log d- \log c}{d-c} = \frac{\sqrt{1- 4k^2}}{2\sqrt{1- 4k^2}-2}\log\left(
\frac{1}{2}+\sqrt{1- 4k^2}/2\right).
\end{equation}
The time $t_*$ gives a time-scale for the transition between the metastable regime
$t<t_*$ during which the system evolves away from equilibrium 
and the relaxation regime $t>t_*$, where the system evolves towards the new
equilibrium state.

\subsection{Explicit computations}

Let us now illustrate the metastability in the BCS-model with a few special
observables: 
using (\ref{lbasisinomegat}) or (\ref{omegatx}) we calculate the expectation 
values in the intermediate states (\ref{omegat}) of some typical 
observables:
\begin{eqnarray} 
\label{x}
&&\omega_t(\sigma^+e^{-i\phi} + \sigma^-e^{i\phi})  =  2\lambda \left( 1 -
2\sin^2(\phi)\left(\lambda^2 e^{-td} + \epsilon^2 e^{-tc}\right)/k^2 \right); 
\\ \label{y}
&&\omega_t(i\sigma^+e^{-i\phi} - i\sigma^-e^{i\phi})  = 
2\lambda \sin(2\phi)e^{-tc}.
\end{eqnarray}
In equations (\ref{x},\ref{y}), the monotone exponential relaxation towards the state
$\omega_{\phi}$ is clear. 
The limits $t \to 0(\infty)$ of the expectation
values give again the equilibrium values in the states
$\omega_{-\phi}(\omega_\phi)$, in particular one computes also:
\begin{eqnarray*}
    \lim_{t \to 0} \omega_t(\sigma^+) & = &\lambda e^{-i\phi} =
  \omega_{-\phi}(\sigma^+);\\
  \lim_{t \to \infty} \omega_t(\sigma^+) & = & \lambda e^{i\phi} =
  \omega_{\phi}(\sigma^+).
\end{eqnarray*}
Note that we have the following bound on 
the time evolution of the `condensate': $\lambda(t) =|\omega_t(\sigma^+)| 
\leq \lambda$, indicating that the `condensation' is suppressed in the 
intermediate states.

The situation is different for the evolution of the expectation value of 
$\sigma^z$, the generator of the gauge group (\ref{ijkgr}):
\begin{equation}\label{z}
\omega_t(\sigma^z)  = -2\epsilon - 4\sin^2(\phi)\frac{\epsilon
\lambda^2}{k^2}\left(e^{-tc} - e^{-td}\right);
\end{equation}
Clearly, this is an example of an observable  invariant under the gauge
transformation group $\mathcal{G}$ (\ref{xinvijk}), and its relaxation is metastable. 
In the time limits \mbox{$t \to 0(\infty)$}, $\omega_t(\sigma^z)$ (\ref{z}) 
tends to the equilibrium value $-2\epsilon$, but 
\mbox{$|\omega_t(\sigma^z)-\omega_{\pm\phi}(\sigma^z)|$} goes through 
a maximum, attained at $t_*$ (\ref{t*}), the exit-time for invariant observables .

\subsection{Temperature Dependence}
To conclude this section about the relaxation behaviour, we give some remarks on
the dependence on the temperature.
At the critical point (\ref{crit}), \ie if $(T \uparrow T_c)$ or $(\beta
\downarrow \beta_c)$, the order parameter $\lambda$ vanishes. 
This also implies the following equalities for $T = T_c$:
\begin{equation*}
  \omega_{-\phi}(X) = \omega_t(X) = \omega_{\phi}(X), \qquad t \geq 0,\ \forall 
  X \in \mathcal{B},
\end{equation*}
and the metastability effects disappears completely.
The constants $(c,d)$, governing the speed of  relaxation decrease to the values $c_c = 2 +
2\cosh(\beta \epsilon)$ resp. $d_c = 4\cosh(\beta\epsilon)$.
Since the expression for the exit-times $t_*(X)$ (\ref{tsterx}) depends
on the ratio of the expectation values $\omega_{-\phi}(a_\phi^0(X))$ and 
$\omega_{-\phi}(a_\phi^+(X)) +\omega_{-\phi}(a_\phi^-(X))$, the behaviour in 
limit $(\lambda \to 0)$ of $t_*(X)$ can be different for different observables 
$X \in M_2^{sa}$. Both $\omega_{-\phi}(a_\phi^0(X))$ and $\omega_{-\phi}(a_\phi^+(X)) +
\omega_{-\phi}(a_\phi^-(X))$ tend to zero at the critical point $(\lambda \to
0)$, but the speed at which they decay to zero can be different. This depend on the
observable $X$ under consideration. 

If we consider the metastable evolution between two groundstates
($T=0,\beta=\infty$), we find that
there are no intermediate states, since the constants $c,d$
(\ref{lbasis}) governing the relaxation speed blow up to infinity as $\beta$
tends to infinity, hence
\begin{equation*}\omega_{t} =
  \left \{ \begin{array}{ll} \omega_{-\phi}  & \quad {\rm for}\ t = 0; \\
 \omega_{\phi}  & \quad {\rm for}\ t > 0. \\
\end{array} \right . 
\end{equation*}
This results from the fact that the dynamical semigroup which we consider
(\ref{gamma}) becomes trivial for groundstates. The detailed analysis of the
dynamics when $T \to 0$ indicates the existence of a non-trivial evolution only
on a different time-scale. 

Hence, if one would like to compare the dynamics at different temperatures, one 
has to rescale the time with an appropriate, temperature dependent, scaling
factor. This scaling and the resulting dynamics ask for an independent analysis
and is out of the scope of our considerations here.

\section{Stability-instability properties of the intermediate states}\label{sectMS}

Here we consider the stability properties of the intermediate states 
$\omega_t(.)$ (\ref{omegat}) on the basis of the correlation inequalities 
(\ref{corp}), \ie we consider the equilibrium or stability conditions:
\begin{equation}\label{corot}
\beta \omega_t(X^*[h_t,X]) \geq
\omega_t(X^*X)\log\left(\frac{\omega_t(X^*X)}{\omega_t(XX^*)}\right),
\end{equation}
where $h_t$ stands for the effective Hamiltonian in the state $\omega_t$:
\begin{equation}\label{ht}
h_t = \epsilon \sigma^z - \omega_t(\sigma^+)\sigma^- -\omega_t(\sigma^-)\sigma^+.
\end{equation}
From section \ref{sectcor} we know that $\omega_t$ is a $\beta$-KMS state if and
only if equation (\ref{corot}) is satisfied for all operators $X \in \mathcal{B}$. 
In this section we analyse to what extend the intermediate states (\ref{omegat})
are still stable, \ie we  check for which operators $X \in M_2$ these
states $\omega_t$ satisfy the inequalities (\ref{corot}).
The meaningful operators in this context are the matrix units in the base
of the spectral decomposition of $h_t$ (\ref{ht}), denoted by $\{E(t)_{++},E(t)_{+-}, E(t)_{-+},E(t)_{--}\}$.
Compute first the expectation values
\begin{equation}\label{omegatEs}
\eqalign{
\omega_t(E(t)_{--}) = \frac{1}{2} +  k_t + \epsilon f_t/(2k_t); \\
\omega_t(E(t)_{-+}) = - \lambda_t f_t/k_t, \\
}
\end{equation}
with $\lambda_t = |\omega_t(\sigma^+)|$, $k_t =\sqrt{\epsilon^2
+\lambda_t^2}$,
the positive eigenvalue of $h_t$ (\ref{ht})
and $f_t = 4\sin^2(\phi)\epsilon \lambda^2\left(e^{-tc} - e^{-td}\right)/k^2$.
The expectation values of the other matrix units can be determined from 
the expressions (\ref{omegatEs}), since $\omega_t(E(t)_{++}) = 1 - \omega_t(E(t)_{--})$ and 
$\omega_t(E(t)_{+-}) = \overline{\omega_t(E(t)_{-+})}$.

\subsection{$\omega_t$ is not an invariant state}
A first observation is that the states $\omega_t,\ t >0$ are not invariant
under the Hamiltonian evolution: 
this is easily derived from the expressions (\ref{omegatEs}), 
calculating 
\begin{equation}\label{notinv}
\omega_t([h_t,E(t)_{+-}])= 2 k_t\omega_t(E(t)_{+-})\ne 0.
\end{equation}

\subsection{Stability}
The position $Q_X$ and momentum $P_X$ observables of a normal mode 
are constructed in the usual way, 
by means of the creation and annihilation operators:
\begin{eqnarray}
a_t^+(X) & = & E(t)_{++}XE(t)_{--}; \label{at+}\\
a_t^-(X) & = & E(t)_{--}XE(t)_{++}, \label{at-}
\end{eqnarray}
yielding the following expression for $Q_X$ and $P_X$:
\begin{eqnarray}
Q_X & = & \frac{1}{\sqrt{2}}\left(a_t^+(X)+a_t^-(X)\right);\label{qx}\\
P_X & = & \frac{i}{\sqrt{2}}\left(a_t^+(X)- a_t^-(X)\right).\label{px}
\end{eqnarray}
These observables $(Q_x,P_X)$ satisfy the canonical dynamical equations:
\begin{equation*}
\eqalign{
i[h_t,Q_X ] =& 2k_tP_{X}; \\
i[h_t,P_X ] =& -2k_tQ_X,
}
\end{equation*}
justifying the name {\em normal coordinates}.
For the sake of completeness we also give $a_t^0(X)$, the canonical 
constant of motion of $X$:
\begin{equation}\label{at0}
a^{0}_t(X) = E(t)_{++}XE(t)_{++} + E(t)_{--}XE(t)_{--} -
\omega_{t}(X)\mathbf{1}
\end{equation} 
These definitions are analogous to the ones in equations
(\ref{a+}), (\ref{a-}) and, (\ref{a0}) in section \ref{sectPS} with an important 
difference. In section \ref{sectPS}, we need the normal modes for the
asymptotic equilibrium state $\omega_\phi$ (\ref{DB2}), whereas the expressions
(\ref{at+}), (\ref{at-}) and, (\ref{at0}) define the normal modes w.r.t.\ the
intermediate state $\omega_t$.    
We can now formulate the following stability statements:
\begin{itemize}
\item
{\em The correlation inequalities (\ref{corot}) are satisfied for all
linear combinations $aQ_X +bP_X$, with $a,b \in \C$ and $Q_X$ and $P_X$ as in 
(\ref{qx}) resp.\ (\ref{px}).}
\newline
For such an operators the correlation inequalities yield:
\begin{equation*}
\omega_t(E(t)_{--}) \geq \omega_t(E(t)_{++}).
\end{equation*}
Since $k_t +  \epsilon f_t/(2k_t) > 0$ (\ref{omegatEs}),
this condition is satisfied for every intermediate state $\omega_t$.
\item
{\em The constants of motion, e.g.\ 
$a^{0}_t(X)$ (\ref{at0}) satisfy the correlation inequalities.}
\newline
An operator $C \in M_2$ is a constant of motion in $\omega_t$ if it satisfies
$[h_t,C]= 0$, such operators can be written as $C = a E(t)_{++} +  b E(t)_{--}$ 
with $a,b \in \C$. It is easy to check that the correlation inequalities
(\ref{corot}) are trivially satisfied. Moreover, this stability property 
holds for any operator $C \in M_2$ satisfying $[h_\rho,C]=0$ 
in a general symmetric product state  with density matrix $\rho$ and effective
Hamiltonian $h_\rho$, see e.g.\ (\ref{corp}).
\end{itemize}
\subsection{Instability}
The instability of the intermediate states is reflected in the following
statements:
{\em The correlation inequalities are satisfied for the creation operators
$a^{+}_t(X)$ (\ref{at+}), $\forall X \in M_2$, but not for the annihilation 
operators $a^{-}_t(X)$ (\ref{at-})}.
\newline
Substituting the operators $a^{+}_t(X)$ (\ref{at+}) resp.\ $a^{-}_t(X)$ 
(\ref{at-}) for $X$ in the energy-entropy balance inequalities 
(\ref{corot}) yields:
\begin{eqnarray}
\label{cora+}
{\rm if}\ X = a^{+}_t(X),\ &{\rm then}\ &  \frac{\omega_t(E(t)_{--})}
{\omega_t(E(t)_{++})} \leq e^{\beta 2k_t};\\
\label{cora-}
{\rm if}\ X = a^{-}_t(X),\ &{\rm then}\ & 
\frac{\omega_t(E(t)_{--})}{\omega_t(E(t)_{++})} \geq e^{\beta 2k_t},
\end{eqnarray}
\ie $\omega_t$ could only be stable under both $a^{+}_t(X)$ and $a^{-}_t(X)$,
$\forall X \in M_2$, if the equality holds. It follows from the gap-equation (\ref{gap2}) 
that these equalities holds only for equilibrium states. 
The intermediate states are stable under the creation operators (\ref{cora+}),
but not under the annihilation operators (\ref{cora-}). 
This can be seen as follows:
using (\ref{omegatEs}), it is readily computed that for all $t > 0$ 
the strict inequality (\ref{cora+}) holds in the limit $\epsilon \to 0$.
Suppose now that there exists a point in  parameter-space $(t',\epsilon')$, 
with $t' > 0,\ \epsilon' > 0$,
such that in that point the inequality (\ref{cora-})
holds. By continuity of the state in parameter-space, 
we have continuity of the function $f\!:(t,\epsilon) \mapsto \omega_t(E(t)_{--}) - 
e^{\beta 2k_t}\omega_t(E(t)_{++})$. Since $f(t',0) <0$ and $f(t',\epsilon') >
0$, there must
exist a point where this function is zero, this would imply that the
intermediate state at that point is an equilibrium state, 
which cannot be true, see e.g.\ (\ref{notinv}). 
This proves that condition (\ref{cora+}) is always satisfied in the 
intermediate states, while (\ref{cora-}) yields that the stability is violated 
for the annihilation operators $a^{-}_t(X)$.

\section{Outlook} \label{sectOL}
All results in this paper concern the BCS-model, and rely 
very much on the mean-field
character \cite{fannes:1980} of this model: all the states under consideration are chosen 
within the set of symmetric product states,  the extension of these results 
to other mean-field models \cite{lauwers:2001} is straight-forward. 
However, this work should be considered as a prototype model of a scheme which
can be generalised to bona fide interacting systems. The main argument for this
is that metastability is formulated in terms of fluctuation operators and their
dynamics. Relying on the general theory of quantum fluctuations
\cite{goderis:1989,goderis:1990} these fluctuation systems are quasi-free or generalised
free systems, and therefore we are  confident that our results have a 
much wider validity far beyond the model considerations of above. We reserve 
this generalisation to interacting systems for a future contribution.

\section*{References}

\bibliographystyle{unsrt}
\bibliography{metastabBCS}

\end{document}